\documentclass[a4paper,11pt]{article}
\usepackage{pos}
\usepackage{comment}

\title{\textit{XMM-Newton} observations of the
extragalactic microquasar S26 and their implications for PeV cosmic rays}
\ShortTitle{S26 as a Pevatron}

\author*[a, b]{Leandro Abaroa}
\author[a,b]{Gustavo E. Romero}
\author[a,b]{Giulio C. Mancuso}
\author[b]{Florencia N. Rizzo}

\affiliation[a]{Instituto Argentino de Radioastronom\'ia,\\
  (CCT La Plata, CONICET; CICPBA; UNLP), C.C.5, (1894) Villa Elisa, Argentina}

\affiliation[b]{Facultad de Ciencias Astron\'omicas y Geof\'{\i}sicas, Universidad Nacional de La Plata,\\
B1900FWA La Plata, Argentina}

\emailAdd{leandroabaroa@gmail.com}
\emailAdd{gustavo.esteban.romero@gmail.com}
\emailAdd{gmancuso@iar.unlp.edu.ar}
\emailAdd{florencianadine.rizzo@gmail.com}

\abstract{The extragalactic microquasar S26 has the most powerful jets observed in accreting binaries, with a kinetic luminosity of $L_{\rm jet}\sim10^{40}\,{\rm erg\,s^{-1}}$. According to the jet-disk symbiosis model, this implies that the accretion power to the stellar black hole at the core of the system should be very super-Eddington, on the order of $L_{\rm acc}\sim L_{\rm jet}$. However, the observed X-ray flux of this system, measured by the \textit{Chandra} and \textit{XMM-Newton} telescopes, indicates an apparent very sub-Eddington accretion luminosity of $L_{\rm X}\approx 10^{37}\,{\rm erg\,s^{-1}}$, orders of magnitude smaller than the jet power. We present here a preliminary investigation of the relationship between jet and disk power, analyze an X-ray observation of S26 obtained with \textit{XMM-Newton}, and propose an explanation for the emission. We also examine the acceleration and distribution of the particles to discuss the feasibility of microquasars as potential PeVatron sources, exploring their ability to produce cosmic rays with energies of about 1 PeV or higher.}

\FullConference{High Energy Phenomena in Relativistic Outflows VIII (HEPROVIII)\\
 23-26 October, 2023\\
Paris, France\\}

\begin{document}
\maketitle

\section{Introduction}

S26 is a microquasar (MQ) located in the galaxy NGC 7793, at a distance of 3.9 Mpc \citep{2003A&A...404...93K}, whose nebula has a size of $\sim 350 \times 185\,{\rm pc}$ \citep{dopita2012}. According to observations in the optical/X-ray band, \cite{2010Natur.466..209P} showed that this MQ has the most powerful jets in accreting binaries, with a mechanical luminosity of $L_{\rm jet}\sim 5\times 10^{40}\,{\rm erg\,s^{-1}}$, and identified nonthermal X-rays produced in the core of the system. On the other hand, \cite{soria2010} resolved the radio lobe structure, and suggested that the radio emission from the terminal region of the jets is consistent with synchrotron radiation. They also analyzed X-ray observations from \textit{Chandra} space telescope, identifying X-ray hotspots $\approx 20\,{\rm pc}$ farther out than the peak of the radio intensity in the lobes, and argued that this emission is most likely of thermal origin. 

A key parameter to characterize MQs is the accretion rate of matter onto the compact object, which proceeds in three basic regimes, depending on the relation of the actual accretion rate to the Eddington rate. In super-Eddington regimes, the inner part of the disk launches powerful winds with mass-loss rates similar to the accretion rate \citep{2023A&A...671A...9A}. Following the jet-disk symbiosis model \citep{1995A&A...293..665F}, the kinetic luminosity of the jet of S26 estimated by \cite{2010Natur.466..209P} implies that the accretion luminosity onto the compact object should be highly super-Eddington, of the order of $L_{\rm acc}\sim L_{\rm jet}$. However, the observed X-ray flux from the core of the system, measured by \textit{Chandra} and \textit{XMM-Newton} telescopes, indicates an apparent very sub-Eddington luminosity of $L_{\rm X}\sim 10^{37}\,{\rm erg\,s^{-1}}$, orders of magnitude smaller than the jet power. 

In this preliminary work, we hypothesize that the compact object is a black hole of 10 solar masses, accreting at super-Eddington rates, and that its actual accretion power is a few times $10^{40}\,{\rm erg\, s^{-1}}$ (the Eddington luminosity for this black hole  is $\sim 10^{39}\,{\rm erg\, s^{-1}}$). We suggest that some energy should also be extracted from the ergosphere of the spinning black hole to provide enough power for the jet \citep{1977MNRAS.179..433B}. The emission from the disk is reprocessed in the dense wind of the disk and is not observable in X-rays. We first analyze the \textit{XMM-Newton} observation and then describe our theoretical model for particle acceleration and distribution. We present our results in Sect. 4 and conclude with some final remarks.

\section{X-ray observation}

We analyzed the \textit{XMM-Newton} data from obsID 0748390901 obtained on December 10, 2014. The observation was performed in the
full frame mode. The data reduction was done using the Science Analysis System (\textsc{SAS}) software package version 20.0.0 and the High Energy Astrophysics Software (\textsc{HEASoft}).
%
We calibrated, filtered, and cleaned the event lists by applying the tasks \texttt{epproc} and \texttt{emproc}, for PN and MOS respectively, using the most up-to-date calibration files. 
We then identified and removed periods dominated by high particle background, defined as a count rate greater than 0.35 ${\rm cts\,s^{-1}}$ for MOS and 0.4 ${\rm cts\,s^{-1}}$ for PN.

We selected a circular region of radius 15 arcsec (which corresponds to a physical distance of $\sim$ 284 pc at the source), centered on $\alpha$=23$^{\rm h}$58$^{\rm m}$00$^{\rm s}$, $\delta$=--32°33'21'' to extract the source spectrum. To account for the background emission, photons were selected from a source-free 30 arcsec circular region on the same chip as the source.
We used the task \texttt{evselect} to retain source events for spectra with FLAG == 0 and single and double events with PATTERN <= 4 for the PN camera, and PATTERN <= 12 for the MOS cameras.
We then generated the response matrix file (rmf) and ancillary response file (arf) using the tasks \texttt{rmfgen} and \texttt{arfgen}, respectively.

\begin{figure}
    \centering
    \includegraphics[width=7cm]{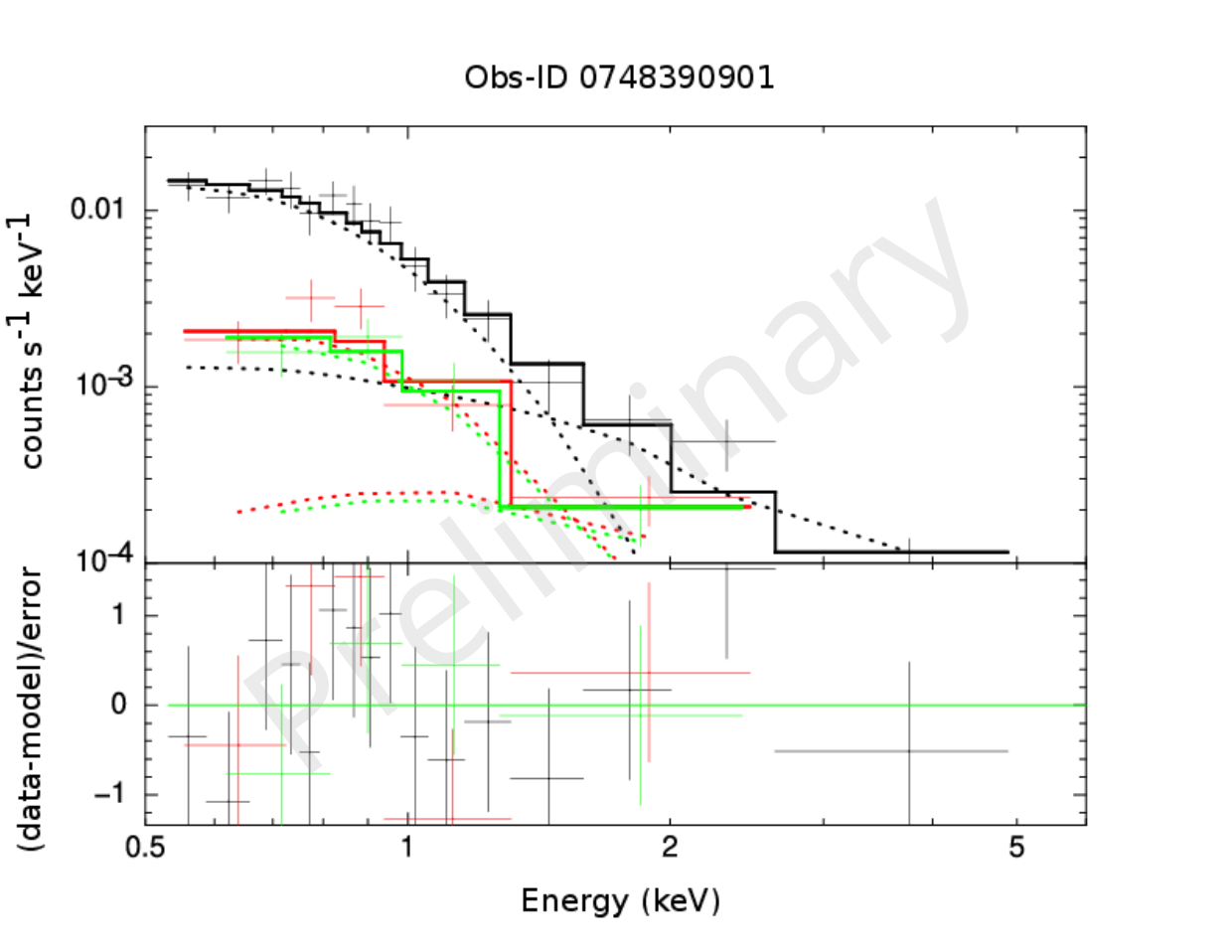}
    \includegraphics[width=8cm]{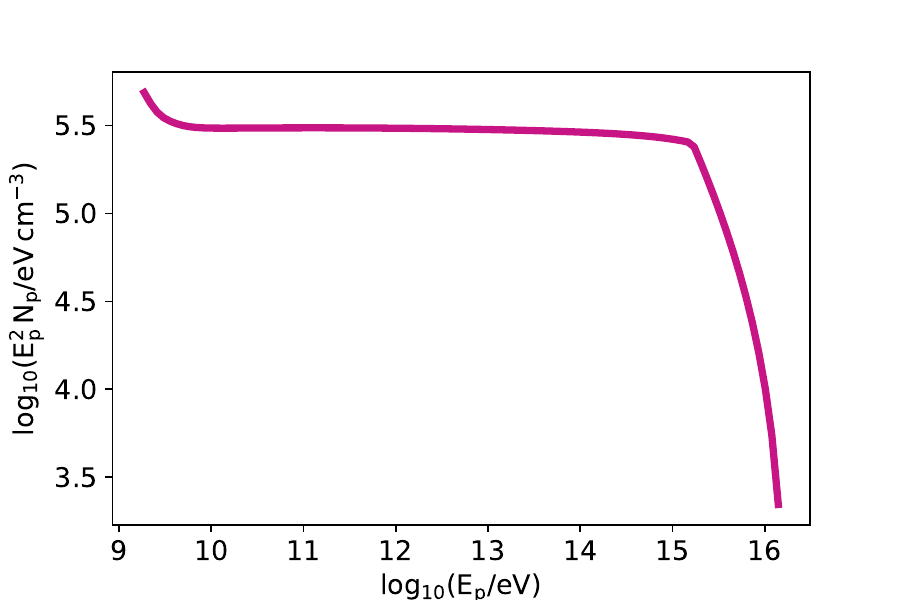}    
    \caption{Left: XSPEC \textsc{constant*tbabs*(bbody+powerlaw)} fitting model of the spectrum of S26 using \textit{XMM-Newton} observation 0748390901. The black, red, and green colors represent the PN, MOS1, and MOS2 cameras, respectively. Right: power-law distribution of protons that are accelerated in the lobes and escape from the system. Their maximum energy is of the order of 10 PeV.}
    \label{fig:fig1}
\end{figure}

We performed the spectral analysis using the \textsc{XSPEC} software v.20.0.0 in the 0.5--6.0 keV energy range. We binned the spectra to have a minimum of 16 counts per bin to allow for Gaussian statistics in the fit using the \texttt{specgroup} tool. We applied a model consisting of a power-law and a blackbody component affected by photoelectric absorption (\texttt{const*tbabs*(powerlaw + bbody}) in XSPEC). We linked the model parameters across the spectra except for $N_{\rm H}$, which was frozen to $1.2 \times 10^{20}$ cm$^{-2}$, and the constant factor, which was fixed to one for PN but allowed to vary freely for the MOS cameras. The spectrum is shown in Fig.~\ref{fig:fig1} (left plot). This model gave a good fit to the observed emission, with $\chi^{2} = 16.44$ for 19 degrees of freedom.
Overall, the parameters obtained with the PN and MOS cameras are consistent with each other within errors. The model fit yielded a power-law photon index $\Gamma$ and a blackbody temperature $kT_{\rm bb}$ which are consistent with the values reported in previous X-ray studies \citep{soria2010}. 
We obtained a 0.5--6.0 keV unabsorbed flux of $1.9^{+0.2}_{-0.3} \times 10^{-14}$ $\rm erg\, cm^{-2}\, s^{-1}$, which corresponds to an unabsorbed luminosity of $\sim 10^{37}$ $\rm erg\, s^{-1}$ for a distance of $d = 3.9$ Mpc \citep{2003A&A...404...93K}.

Our model suggests that the emission from S26 could have two possible origins, one thermal and one non-thermal. This is consistent with the results reported by \cite{soria2010}, who showed that the X-ray spectrum of S26 can be modeled by a combination of thermal and nonthermal components, where the former is associated with the lobes and the latter with the emission from the core.

\section{Disk and jet models}

We assume that the compact object is a black hole (BH) of mass $M_{\rm BH}=10\,M_{\odot}$ and accretes matter at a super-Eddington rate at the outer part of the disk, $\dot{m}=\dot{M}_{\rm input}/\dot{M}_{\rm Edd} \gg 1$, where $\dot{M}_{\rm input}$ is the input of mass per unit of time. The Eddington rate is given by $\Dot{M}_{\rm{Edd}}= L_{\rm{Edd}}/\eta c^2 \approx 2.2\times 10^{-8} M_{\rm BH} \, {\rm yr^{-1}} = 1.4 \times 10^{18} M M_\odot^{-1} \, \rm{g \, s^{-1}}$,
with $L_{\rm Edd}$ the Eddington luminosity (defined as the luminosity required to balance the attractive gravitational pull of the BH by radiation pressure), $\eta \approx 0.1$ the accretion efficiency, and $c$ the speed of light. The self-regulation of the inner accretion disk to the Eddington rate results in a total mass-loss rate in winds $\dot{M}_{\rm w}$ that is approximately equal to the accretion input, $\dot{M}_{\rm w}\approx \dot{M}_{\rm input}\gg \dot{M}_{\rm Edd}$. In this supercritical scenario, the disk will be completely covered by the opaque wind (whose photosphere extends to $z_{\rm photo}\approx 10^{11}\,{\rm cm}$, e.g., \cite{2023arXiv231115050A}), so the radiation of the disk can not reach the observer because it is absorbed and reprocessed in the wind and shifted to lower energies. The accretion power at a radius $r_{\rm d}$ is given by $L_{\rm acc}(r_{\rm d})\approx \dot{M}_{\rm acc}(r_{\rm d})\,c^2$, where $\dot{M}_{\rm acc}(r_{\rm d})=\dot{M}_{\rm input}r_{\rm d}/r_{\rm crit}$, with $r_{\rm crit}\approx 40\dot{m}r_{\rm g}$ the critical radius. This radius corresponds to the distance to the BH at which the outer standard disk changes to the inner disk dominated by radiation.

Two jets are generated near the BH in opposite directions, perpendicular to the orbital plane of the binary system. Following the disk-jet coupling hypothesis of  
\cite{1995A&A...293..665F}, the kinetic power of each jet, $L_{\rm j}$, is assumed to be proportional to the accretion power at the magnetically-saturated radius, $r_{\rm ms}$ (where the poloidal magnetic flux becomes dynamically important), $L_{\rm j}=q_{\rm j}\,L_{\rm acc}(r_{\rm ms})$, with $q_{\rm j}$ the efficiency in the transfer of power from the disk to the jet. We assume that the disk is saturated at a distance from the black hole of $r_{\rm ms}=100\,r_{\rm g}$. At this radius, the accretion rate is $\dot{M}_{\rm acc}(r_{\rm ms})=2.5\dot{M}_{\rm Edd}=3.5\times10^{19}\,{\rm g\, s^{-1}}$, so the accretion power is $L_{\rm acc}(r_{\rm ms})=3.15\times10^{40}\,{\rm erg\, s^{-1}}$. This luminosity is not enough to explain the jet power in S26; therefore, some energy should also be extracted from the ergosphere of the spinning black hole \citep{1977MNRAS.179..433B}. Considering that the jet kinetic luminosity of S26 estimated by \cite{2010Natur.466..209P} is $L_{\rm j}=5\times10^{40}\,{\rm erg\, s^{-1}}$, then the efficiency factor in the transformation of accretion into jet power should be $q_{\rm j}=L_{\rm j}/L_{\rm acc}\approx 1.58$.

We assume that the jets are launched at a distance $z_0=100\,r_{\rm g}$ from the BH with a typical semi-opening angle of $\theta_{\rm j}\approx 5.7^{\circ}$, so the radius-to-height relation at any point is estimated to be $r_{\rm j}\approx0.1\,z_{\rm j}$. A fraction $q_{\rm rel}=0.1$ of the jet power is in the form of relativistic particles, $L_{\rm rel}=q_{\rm rel}\,L_{\rm j}$, and we include both hadronic and leptonic content, $L_{\rm rel}=L_{\rm p}+L_{\rm e}$. We assume a hadronic-dominant scenario at the terminal regions of the jet, $L_{\rm p}=100\,L_{\rm e}$. 


Since nonthermal X-rays are detected from the core of S26, recollimation shocks should take place near the base of the jet, but above the photosphere of the disk-driven wind to account for the emission, $z_{\rm recoll}>z_{\rm photo}$. From an order-of-magnitude analysis of the available power, this jet-disk model may explain the core X-ray luminosity inferred from observations $(10^{37}\,{\rm erg\, s^{-1}})$ and the jet power $(10^{40}\,{\rm erg\, s^{-1}})$. The full calculations of the nonthermal processes will be presented in a forthcoming paper. 

At the head of the jet two more shocks are generated: a forward radiative shock that propagates into the interstellar medium (ISM), and a reverse adiabatic shock that propagates to the jet and forms the thermal cocoon. In adiabatic shocks, particles can be accelerated to relativistic energies and produce nonthermal emission. In radiative shocks, the gas emits large amounts of thermal radiation. The radii of the shock regions are taken from \cite{2010Natur.466..209P} (we refer to section 3.3 of \cite{2023A&A...671A...9A} for characterization of shocks). In what follows, we concentrate only on the backward shock region at the head of the jets --where the radio hotspots were identified by \cite{soria2010}-- because we are interested in MQs as cosmic ray generators.

\section{Particle acceleration and distribution}
Particles accelerated in the reverse shocks at jet terminal regions of the jet can cool through different processes and produce nonthermal radiation. The diffusive acceleration rate of the particles is given by  $t^{-1}_{\rm{acc}}=\eta_{\rm acc}~e~Z~c~B/E$, where $e$ is the electric charge, $Z$ the atomic number, $c$ the speed of light, $B=10\,\mu{\rm G}$ the magnetic field (which corresponds to that of the compressed ISM), $E$ is the energy of the particle, and $\eta_{\rm acc}=10^{-1}$ the acceleration efficiency. 
The maximum energy for each kind of particle can be inferred by looking at the point where the acceleration rate is equal to the total cooling or escape rate. 

Our model is lepto-hadronic, and so we calculate the following cooling processes numerically: synchrotron (interaction of protons and electrons with the ambient magnetic field), inverse Compton (IC, collision of relativistic electrons with photons of the ambient radiation field), Bremsstrahlung (Coulombian interactions between relativistic electrons and cold matter), photo-hadronic interactions (interaction of highly relativistic protons with photons of the ambient radiation field), and proton-proton (collision of relativistic protons with cold matter). The number density of cold protons is $n_{\rm p}\approx 0.7\,{\rm cm^{-3}}$ \citep{2010Natur.466..209P}.

We then investigate the evolution of particles that are accelerated at the reverse shocks in the jet lobes and injected into the ISM. The relativistic particles have a distribution given by ${\rm d}N= n(\vec{r},E,t){\rm d}E{\rm d}V$, where $n$ is the number density of particles, $t$ the time, $\vec{r}$ the position, $V$ the volume, and $E$ the energy. The evolution of this distribution is determined by the transport equation \citep[see e.g.,][]{1964ocr..book.....G}. We solve this equation numerically in steady state and in the one-zone approximation: $    \partial/\partial E [{\rm d}E N(E)/{{\rm d}t}]+N(E)/t_{\rm esc}=Q(E)$, where $t_{\rm esc}$ is the escape time, and the particle injection function, $Q(E)=Q_{0}E^{-p}\exp{(-E/E_{\rm max})}$ is a power-law in the energy with an exponential cutoff and a spectral index $p=2$, which is characteristic of the Fermi first-order acceleration mechanism.

Timescales of particles of the reverse shock in the lobes show synchrotron as the dominant cooling mechanism for leptons, whereas proton-proton interaction is the dominant cooling mechanism for hadrons. The power-law distribution of protons is shown in Fig. \ref{fig:fig1} (right plot). Electrons reach a maximum energy of 5 PeV, while protons are accelerated up to  10 PeV energies. These ultra-relativistic protons will be injected in the ISM and will in turn spread through the intergalactic medium, populating with cosmic rays the whole galaxy.

\section{Conclusions}

We have reduced and analyzed \textit{XMM-Newton} data of the microquasar S26 and found that the X-ray emission fits well with a two-component model: black body (soft emission) plus power-law (hard emission). This is in agreement with previous analysis of X-ray observations  \citep{2010Natur.466..209P,soria2010}. The detected nonthermal component of the X-ray radiation could have originated from synchrotron emission from the core, produced by electrons accelerated at the base of the jet, whereas the thermal component could be explained with radiation produced by non-relativistic Bremsstrahlung in the forward shock at the jet-lobes.

We have also proposed that the three-order-of-magnitude difference between the jet and apparent disk luminosities in S26 is caused by the complete absorption of the disk radiation by the wind ejected from the super-Eddington disk. Moreover, we suggest that some of the energy to power the jet must be extracted from the ergosphere of the spinning black hole. The observed soft X-ray flux should be then produced near the base of the jet but above the wind photosphere.

One of the main results of this preliminary work is that protons accelerated downstream in the lobes reach energies close to 10 PeV and are cooled by $pp$ interactions. These ultra-relativistic protons will in turn spread through the intergalactic medium, populating with cosmic rays the whole galaxy. Super-Eddington MQs could be thus considered as potential PeVatron sources. 

The cooling timescales, the spectral energy distributions for each radiative process involved in this system -- and their comparison with the observed data --, as well as a more detailed examination of S26 will be presented in a near-future work.

\acknowledgments{LA acknowledges the UNLP and the HEPRO  VIII organizers. GER was funded by  PID2022-136828NB-C41/AEI/10.13039/501100011033/ and through the ``Unit of Excellence María de Maeztu 2020-2023'' award to the Institute of Cosmos Sciences (CEX2019-000918-M). Additional support came from PIP 0554 (CONICET).}

\bibliographystyle{jhep}
\bibliography{main}

\end{document}